\documentclass[12pt]{article}
\usepackage{amssymb}
\usepackage{amsmath}
\usepackage[space]{cite}
\usepackage{algorithm}
\usepackage{bbm}
\usepackage[noend]{algpseudocode}
\usepackage[normalem]{ulem}
\usepackage{url}
\usepackage{graphicx}
\usepackage{adjustbox}
\usepackage{caption}
\usepackage{subcaption}
\usepackage{enumitem}   

\usepackage{MnSymbol}

\usepackage{cases}

\numberwithin{equation}{section}

\newcommand{\non}{\nonumber}
\newcommand{\id}{\mathbb{I}}

\newcommand{\diag}{\mathop{\rm diag}\nolimits}

\newcommand{\cut}[1]{\ifmmode\text{\textcolor{red}{\sout{\ensuremath{#1}}}}\else\textcolor{red}{\sout{#1}}\fi}

\newcommand{\RN}[1]{%
  \textup{\uppercase\expandafter{\romannumeral#1}}%
}

\usepackage{quantikz}
\usetikzlibrary{circuits.logic.US}
\tikzset{
    gateO/.style={
        draw,
        circle,
        minimum width=0.5em,
        inner sep=2pt    }
}
\DeclareExpandableDocumentCommand{\gateO}{O{}{m}}{|[gateO,#1]| {#2} \qw}

\tikzset{
    gateOS/.style={
        draw,
        circle,
        minimum width=0.5em,
        inner sep=2pt,
		fill=red!20}
}
\DeclareExpandableDocumentCommand{\gateOS}{O{}{m}}{|[gateOS,#1]| {#2} \qw}

\begin{document}

\begin{titlepage}
\vspace{.5in}
\begin{center}

{\LARGE Spin-$s$ Dicke states and their preparation}\\
\vspace{1in}
\large Rafael I. Nepomechie\footnote{
Physics Department, P.O. Box 248046, University of Miami,
Coral Gables, FL 33124 USA}${}^{,}$\footnote{{\tt nepomechie@miami.edu}},
Francesco Ravanini\footnote{
Dept. of Physics and Astronomy, University of Bologna,
Via Irnerio 46, I-40126 Bologna, Italy  
{\tt francesco.ravanini@unibo.it}}${}^{,}$\footnote{INFN, Sezione di Bologna, Via Irnerio 46, I-40126 Bologna, Italy} and
David Raveh${}^{1,}$\footnote{{\tt dxr921@miami.edu}}\\[0.2in] 
\end{center}

\vspace{.5in}

\begin{abstract}
We introduce the notion of $su(2)$ spin-$s$ Dicke states, which are higher-spin generalizations of usual (spin-1/2) Dicke states.
These multi-qudit states can be expressed as superpositions of $su(2s+1)$ qudit Dicke states. They satisfy a recursion formula, which we use to formulate an efficient quantum circuit for their preparation, whose size scales as $sk(2sn-k)$, where $n$ is the number of qudits and $k$ is the number of times the total spin-lowering operator is applied to the highest-weight state.
The algorithm is deterministic and does not require ancillary qudits.
\end{abstract}

\end{titlepage}

\setcounter{footnote}{0}

\section{Introduction}\label{sec:intro}

Quantum state preparation is a fundamental task in quantum computing \cite{Nielsen:2019}. The cost of preparing a general quantum state scales exponentially with the number of qubits (or qudits, for $d$-level systems),
see e.g. \cite{Barenco:1995, Kaye:2004, Mottonen:2004, Shende:2006, Plesch:2011}. Hence, quantum states that can be prepared efficiently are of particular interest. Dicke states \cite{Dicke:1954zz} constitute one such example. These states, which we denote here by $|D_{n,k}\rangle$, are completely symmetric $n$-qubit states of $k$ $|1\rangle$'s and $n-k$ $|0\rangle$'s, for instance
\begin{equation}
    |D_{3,2}\rangle = \frac{1}{\sqrt{3}}\left(|0 1 1 \rangle + |1 0 1  \rangle +|1 1 0 \rangle \right) \,,
\end{equation}
where the tensor product is understood, e.g. 
$|011\rangle = |0\rangle \otimes |1\rangle \otimes |1\rangle$. Such states have numerous applications, including quantum networking \cite{Prevedel:2009}, quantum metrology \cite{Toth:2012}, optimization \cite{Farhi:2014}, and quantum compression \cite{Bartschi2019}. An efficient algorithm for preparing Dicke states was given in \cite{Bartschi2019}, see also \cite{Chakraborty:2012, Bartschi:2022}. This construction was used
recently as the starting point for preparing exact eigenstates of the Heisenberg spin chain \cite{VanDyke:2021kvq, VanDyke:2021nuz, Li:2022czv} via coordinate Bethe ansatz \cite{Bethe:1931hc, Gaudin:1983}.

A generalization of Dicke states $|D_{n,k}\rangle$ to higher-level systems is given by qudit Dicke states, which are multi-qudit completely symmetric basis states (a precise definition can be found in Appendix \ref{sec:relating}); and an algorithm for preparing these states, generalizing \cite{Bartschi2019}, was given in \cite{Nepomechie:2023lge}.

Additional types of quantum states that can be prepared efficiently include
the $q$-deformation of qubit \cite{Li:2015} and qudit \cite{Raveh:2023iyy} Dicke states, uniform and cyclic quantum states \cite{Mozafari:2022} and W states \cite{Yeh:2023}.

In this paper we introduce the notion of \emph{higher-spin} Dicke states,
and we formulate a deterministic algorithm for preparing these states that does not require ancillas. 
These multi-qudit states differ from the above-mentioned qudit Dicke states, and can in fact be prepared with significantly simpler circuits. We expect that these states may be useful for generalizing the many 
applications of (qubit) Dicke states to qudits, and may serve as the 
starting point for preparing exact eigenstates of integrable higher-spin Heisenberg chains via coordinate Bethe ansatz \cite{Crampe:2011}.

Specifically, we consider qudits with dimension $d=2s+1$, where $s= 1/2\,, 1\,, 3/2\,, \ldots$, corresponding to spin-$s$ spins. We denote the basis by
\begin{equation}
    |0\rangle = \begin{pmatrix}
    1\\
    0\\
    \vdots\\
    0\end{pmatrix} \,, 
|1\rangle = \begin{pmatrix}
    0\\
    1\\
    \vdots\\
    0\end{pmatrix} \,, \ldots \,,
|2s\rangle = \begin{pmatrix}
    0\\
    0\\
    \vdots\\
    1\end{pmatrix} \,,
\end{equation}
as usual. The total spin operators $\vec{\mathbb{S}}$ are given by 
\begin{equation}
    \vec{\mathbb{S}} = \sum_{i=0}^{n-1} \vec{S}_i \,, \qquad
    \vec{S}_i = 
    \stackrel{\stackrel{n-1}{\downarrow}}{\id}
    \otimes \ldots \otimes \id \otimes 
    \stackrel{\stackrel{i}{\downarrow}}{\vec{S}}
    \otimes \id \otimes \ldots \otimes 
    \stackrel{\stackrel{0}{\downarrow}}{\id} \,,
    \label{bbS}
\end{equation}
where $\vec{S} = (S^x\,, S^y\,, S^z)$ are $(2s+1) \times (2s+1)$ matrices that obey the $su(2)$ algebra $\left[ S^x\,, S^y \right] = i S^z$, etc.,
and $\id$ is the  $(2s+1) \times (2s+1)$ identity matrix.
As usual, we take $S^z$ to be the diagonal matrix
\begin{equation}
S^z = \diag\left(s\,, s-1\,, \ldots \,, -(s-1) \,, -s \right) \,.
\end{equation}

For a system of $n$ such qudits, we define spin-$s$ Dicke states $|D^{(s)}_{n,k}\rangle$ as the states obtained by applying $k$ times
the total spin-lowering operator $\mathbb{S}^-$
on the highest-weight state.
More precisely, 
\begin{equation}
    |D^{(s)}_{n,k}\rangle = a^{(s)}_{n,k}  \left(\mathbb{S}^- \right)^k |0\rangle^{\otimes n} \,, \qquad k = 0, 1, \ldots, 2sn \,, 
    \label{spinsDicke}
\end{equation}
where $\mathbb{S}^- = \mathbb{S}^x - i \mathbb{S}^y$ is the total spin-lowering operator, $|0\rangle^{\otimes n}$ is the state with all $n$ spins ``up'' (with $\mathbb{S}^z$-eigenvalue $\mathbbm{m} = s n$), and $a^{(s)}_{n,k}$ is the normalization factor\footnote{This factor can be derived using the familiar fact $$\mathbb{S}^-\,|\mathbbm{s},\mathbbm{m}\rangle = \sqrt{(\mathbbm{s}+\mathbbm{m})(\mathbbm{s}+1-\mathbbm{m})}\, |\mathbbm{s},\mathbbm{m}-1\rangle \,, $$
where $|\mathbbm{s},\mathbbm{m}\rangle$ are simultaneous eigenstates of $\vec{\mathbb{S}}^{\, 2}$ and $\mathbb{S}^z$,
and the fact that here $\mathbbm{s}=s n$.}
\begin{equation}
    a^{(s)}_{n,k} = \frac{1}{k!\,\sqrt{\binom{2sn}{k}}} \,.
    \label{normalization}
\end{equation}
These states are exact ground states of ferromagnetic spin-$s$ 
Heisenberg Hamiltonians, 
and of a spin-$s$ version of the Lipkin-Meshkov-Glick  
\cite{Lipkin:1964yk} Hamiltonian
$-\vec{\mathbb{S}}^{\, 2} = -\sum_{i,j} \vec{S}_i \cdot \vec{S}_j$.

The spin-$s$ Dicke states take the closed-form expression
\begin{equation}
|D^{(s)}_{n,k}\rangle= \sum_{\substack{j_i=0,1,\dots,2s\\ j_0+j_1+\dots+j_{n-1}=k}}
\sqrt{\frac{
\binom{2s}{j_0}
\binom{2s}{j_1}\dots
\binom{2s}{j_{n-1}}
}{
\binom{2sn}{k}}}\,|j_{n-1}\dots j_1 j_0\rangle\,,
\label{closed}
\end{equation}
see Appendix \ref{sec:relating} for a complete proof of this fact. Thus, for $s=1/2$ the spin-$s$ Dicke states reduce to the usual Dicke states, i.e. $|D^{(1/2)}_{n,k}\rangle = |D_{n,k}\rangle$. Further, for $s>1/2$, these states can be expressed as linear combinations of $(2s+1)$-level qudit Dicke states. A simple example with $s=1$ is the state
\begin{equation}
    |D^{(1)}_{3, 2}\rangle = \frac{2}{\sqrt{15}}\left(|0 1 1 \rangle + |1 0 1  \rangle +|1 1 0 \rangle \right) +  \frac{1}{\sqrt{15}}\left(|0 0 2 \rangle + |0 2 0  \rangle +|2 0 0 \rangle \right) \,.
    \label{example}
\end{equation}
The general relation between higher-spin and qudit Dicke states is given by \eqref{decomp}, \eqref{alphas}. Higher-spin Dicke states are entangled, and we include here a computation of their bipartite entanglement entropy.
We remark that these states have the ``duality'' (charge conjugation) transformation property
\begin{equation}
    \mathcal{C}^{\otimes n}\,  |D^{(s)}_{n,k}\rangle =  
    |D^{(s)}_{n, 2 s n - k}\rangle \,, \qquad  
    \mathcal{C} = \begin{pmatrix}
    & &  1 \\
    & \udots & \\
    1 & & 
    \end{pmatrix} \,,
    \label{duality}
\end{equation}
which maps $k \mapsto 2sn-k$.
To our knowledge, such higher-spin Dicke states have not heretofore been systematically studied.\footnote{After this work was completed, we became aware of \cite{Liu:2015}, which has some overlap with Appendix \ref{sec:relating}. We thank a referee for bringing this reference to our attention.}

An outline of the reminder of the paper is as follows. In Section \ref{sec:general}, we present a recursive construction of higher-spin Dicke states on a quantum computer. (Of course, the construction \eqref{spinsDicke} cannot be directly implemented on a quantum computer, since the total spin-lowering operator $\mathbb{S}^-$ is not unitary.)
The key idea is that, as for the case of usual Dicke states \cite{Bartschi2019} and qudit Dicke states \cite{Nepomechie:2023lge, Raveh:2023iyy}, these states satisfy a recursion formula \eqref{recursion}, which is proved in Appendix \ref{sec:proof}. The  reference state \eqref{refstate}, whose choice requires considerable care, also plays an important role in this construction. The problem reduces to finding explicit circuits for certain operators $T$. As a warm-up for determining these $T$ operators, we briefly review in Section \ref{sec:spin12} the case $s=1/2$ \cite{Bartschi2019, Nepomechie:2023lge}. In Section \ref{sec:spin1} we consider the case $s=1$; and we finally treat the general spin-$s$ case in Section \ref{sec:spins}.  We conclude with a brief discussion in Section \ref{sec:discuss}. We complement the paper with some appendices; in particular, Appendix \ref{sec:relating} shows the relation between higher-spin and qudit Dicke states, and Appendix \ref{sec:EE} contains the computation of the entanglement entropy. Code in cirq \cite{cirq} for simulating the circuits presented here is provided in the Supplementary Material.

\section{Recursive construction}\label{sec:general} 

We assume that a spin-$s$ Dicke state \eqref{spinsDicke} can be generated by a unitary operator $U^{(s)}_n$ acting on a simple ``reference'' state $|\psi^{(s)}_{n,k}\rangle$
\begin{equation}
    |D^{(s)}_{n,k}\rangle = U^{(s)}_n\,   |\psi^{(s)}_{n,k}\rangle\,, \qquad 
    k = 0, 1, \ldots, 2sn \,,
    \label{Uop}
\end{equation}
where $U^{(s)}_n$ is independent of $k$. 

In order to specify the reference state $|\psi^{(s)}_{n,k}\rangle$ for a given value of $k$, it is necessary to first define $\ell$ to be the unique integer satisfying 
\begin{equation}
k = 2s\ell + i
     \,, \qquad  0\leq i<2s\,,
    \label{kli1}
\end{equation}
so that 
\begin{equation}
    \ell = \lfloor \frac{k}{2s} \rfloor \in \{0, 1, \ldots, n\} \,, \qquad i=k-2s\ell \in  \{0, 1,  \ldots, 2s-1\} \,,
    \label{kli2}
\end{equation}
where $\lfloor \ldots \rfloor$ denotes floor. 
The reference state $|\psi^{(s)}_{n,k}\rangle$ is then given by the product state
\begin{equation}
    |\psi^{(s)}_{n,k}\rangle \equiv
    |\psi^{(s)\, i}_{n \,; \ell}\rangle = |0\rangle^{\otimes(n-\ell-1)}|i\rangle |2s\rangle^{\otimes \ell} \,.
    \label{refstate}
\end{equation}
These reference states have been engineered so that
they reduce for $n=1$ to basis states $|k\rangle$ 
\begin{equation}
    |\psi^{(s)}_{1,k}\rangle = |k\rangle = |D^{(s)}_{1,k}\rangle\,, \qquad 
    k = 0, 1, \ldots, 2s \,,
\end{equation}
which implies that $U^{(s)}_n$ in \eqref{Uop}
reduces for  $n=1$ to the identity matrix 
\begin{equation}
 U^{(s)}_1 = \id \,.
 \label{init}
\end{equation}

A key feature of spin-$s$ Dicke states \eqref{spinsDicke} is that they obey a recursion formula
\begin{equation}
    |D^{(s)}_{n,k}\rangle = \sum_{j=0}^{2s} c^{(s)}_{n,k,j}\, |D^{(s)}_{n-1,k-j}\rangle \otimes |j\rangle \,,
    \label{recursion}
\end{equation}
with
\begin{equation}
  c^{(s)}_{n,k,j} 
  = \sqrt{\frac{\binom{2s}{j}\binom{2sn-2s}{k-j}}{\binom{2sn}{k}}}  \,,
  \label{ccoefs}
\end{equation}
whose proof is given in Appendix \ref{sec:proof}. 
Similarly to \cite{Bartschi2019, Nepomechie:2023lge, Raveh:2023iyy}, let us now define a unitary operator $W^{(s)}_n$ that implements a corresponding mapping on the reference states
\begin{equation}
    W^{(s)}_n\, |\psi^{(s)}_{n,k}\rangle = \sum_{j=0}^{2s} c^{(s)}_{n,k,j}\, |\psi^{(s)}_{n-1,k-j}\rangle \otimes |j\rangle \,, \qquad n = 2, 3, \ldots \,.
    \label{Wop}
\end{equation}
Note that $W^{(s)}_n$, like $U^{(s)}_n$, is independent of $k$. 
Making use of \eqref{Uop} in both sides of \eqref{recursion}, we see that $U^{(s)}_n$ satisfies the recursion
\begin{equation}
    U^{(s)}_n = \left( U^{(s)}_{n-1} \otimes \id \right)\, W^{(s)}_n \,.
\end{equation}
Telescoping the recursion, and imposing the initial condition \eqref{init}, we conclude that $U^{(s)}_n$ is given by an ordered product of $W$ operators 
\begin{equation}
U^{(s)}_n =  \overset{\curvearrowright}{\prod_{m=2}^{n}} 
\left(W^{(s)}_m \otimes  \id^{\otimes(n-m)} \right)\,,
\label{Uresult}
\end{equation}
where the product goes from left to right with increasing $m$.

The problem of constructing a quantum circuit for $U^{(s)}_n$
therefore reduces to finding circuits for the $W^{(s)}_{m}$ operators. The strategy for accomplishing the latter is to look for operators 
$T^{(s)}_{m,k} \equiv T^{(s)\, i}_{m; \ell}$ (recall the definitions \eqref{kli1} and \eqref{kli2} for $\ell$ and $i$), which \emph{do} depend on $k$, with the following properties
\begin{numcases}{ T^{(s)}_{m,k'}\, |\psi^{(s)}_{m,k}\rangle =}
    |\psi^{(s)}_{m,k}\rangle & for $\quad k'< k$ \label{prop1} \\[0.1in]
    W^{(s)}_m\, |\psi^{(s)}_{m,k}\rangle  & for  $\quad k' = k$
    \label{prop2}
    \end{numcases} 
and 
\begin{equation}
    T^{(s)}_{m,k'}\, \left(T^{(s)}_{m,k}\,  |\psi^{(s)}_{m,k}\rangle \right)
    =  \left(T^{(s)}_{m,k}\,  |\psi^{(s)}_{m,k}\rangle \right) \quad 
    \text{for}\quad k' > k \,,
    \label{prop3}
\end{equation}
where $W^{(s)}_m\, |\psi^{(s)}_{n,k}\rangle$ in \eqref{prop2} is given by \eqref{Wop}. An operator $W^{(s)}_m$ that performs the mapping \eqref{Wop}
is therefore given by an ordered product of all the $T$ operators
\begin{equation}
W^{(s)}_m = 
\overset{\curvearrowleft}{\prod_{k=1}^{2sm-1}}
T^{(s)}_{m,k}	\,, 
\label{Wresult}
\end{equation}
where the product goes from right to left with increasing $k$. 

We see from \eqref{Wresult} that the number of $T$ operators in $W^{(s)}_m$ is $2sm-1$; and from \eqref{Uresult} we conclude that
the number of $T$ operators in $U^{(s)}_n$ is
\begin{equation}
    \sum_{m=2}^n (2sm-1)  = \mathcal{O}(n^2 s) \,.
    \label{Tcount}
\end{equation}
However, we shall later argue that the number of $T$ operators can be reduced, see \eqref{Tcounts}.

To summarize, spin-$s$ Dicke states \eqref{spinsDicke} are generated by \eqref{Uop}, where the reference state $|\psi^{(s)}_{n,k}\rangle$ is given by \eqref{refstate}, the unitary operator $U^{(s)}_n$ is given in terms of $W$'s by \eqref{Uresult}, and the $W$'s are given in terms of $T$'s by \eqref{Wresult}. It remains to find explicit circuits for the $T$'s, to which the reminder of this paper is largely dedicated. As a warm-up, we begin by reviewing the case $s=1/2$ in Sec. \ref{sec:spin12}; we then treat the case $s=1$ in Sec. \ref{sec:spin1}, and we finally consider the general spin-$s$ case in Sec. \ref{sec:spins}.

\section{The case $s=1/2$}\label{sec:spin12}

\begin{figure}[htb]
	\centering
    \begin{subfigure}{0.5\textwidth}
        \centering
\begin{adjustbox}{width=0.5\textwidth}
\begin{quantikz}
\lstick{$0$} & \ctrl{3}  &  \gate{R(\theta_1)} \vqw{2} & \ctrl{3}  & \qw \\
\vdots \\
\lstick{$k-1$} & \qw  & \ctrl{0}  & \qw  & \qw  \\
\lstick{$k$} & \targ{}  &  \ctrl{-1}  & \targ{}   & \qw \\
\vdots\\
\lstick{$m-1$}&\qw &\qw &\qw &\qw
\end{quantikz}
\end{adjustbox}
\caption{$T^{(1/2)}_{m,k}$ with $k>1$}
\label{fig:Il}
	 \end{subfigure}%
	\begin{subfigure}{0.5\textwidth}
      \centering
\begin{adjustbox}{width=0.5\textwidth, raise=5.5em}
\begin{quantikz}
\lstick{$0$} & \ctrl{1}  &  \gate{R(\theta_1)} \vqw{1} & 
\ctrl{1}  & \qw \\
\lstick{$k=1$} & \targ{} &  \ctrl{-1} & 
\targ{}   & \qw \\
\vdots \\
\lstick{$m-1$}&\qw&\qw&\qw&\qw
\end{quantikz}
\end{adjustbox}
\caption{$T^{(1/2)}_{m,k}$ with $k=1$}
\label{fig:I1}
    \end{subfigure}
\caption{Circuit diagrams for $T^{(1/2)}_{m,k}$}
\label{fig:T12ops}
\end{figure}
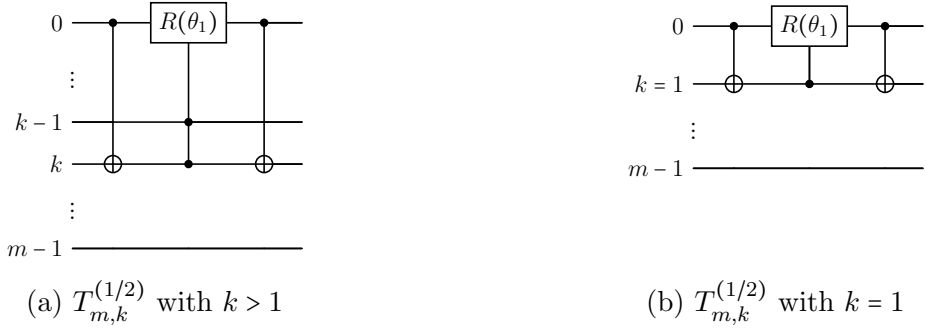

For the case $s=1/2$, which corresponds to usual qubit Dicke states,
we see from \eqref{kli1} that $\ell=k$ and $i=0$; hence, the reference state \eqref{refstate} with $n=m$ reduces to 
\begin{equation}
    |\psi^{(1/2)}_{m,k}\rangle  = |0\rangle^{\otimes(m-k)} |1\rangle^{\otimes k} \,.
\end{equation}
The action of the $W$ operator on this state is given by \eqref{Wop} 
\begin{equation}
    W^{(1/2)}_m\, |0\rangle^{\otimes(m-k)} |1\rangle^{\otimes k}
    =  c^{(1/2)}_{m,k,0}\, 
    |0\rangle^{\otimes(m-k-1)} |1\rangle^{\otimes k}|0\rangle
    + c^{(1/2)}_{m,k,1}\,  |0\rangle^{\otimes(m-k)} |1\rangle^{\otimes k}
    \,. 
    \label{Wop12}
\end{equation}
We define a 3-qubit operator $T^{(1/2)}_{m,k}$ (denoted by $\RN{1}_{m,k}$ in \cite{Nepomechie:2023lge, Raveh:2023iyy}) that performs the mapping \eqref{prop2} with \eqref{Wop12}, which acts on the $k$th, $(k-1)$th, and $0$th qubit, as follows
\begin{equation}
T^{(1/2)}_{m,k}:\quad |0\rangle_k\, |1\rangle_{k-1}\, 
|1\rangle_0 \mapsto
c^{(1/2)}_{m,k,0}\, 
|1\rangle_k\, |1\rangle_{k-1}\, 
|0\rangle_0 
+
c^{(1/2)}_{m,k,1}\, 
|0\rangle_k\, 
|1\rangle_{k-1}\, 
|1\rangle_0
\,,
\label{T12action}
\end{equation}
and otherwise acts as identity (as long as the $0$th qubit is in the state $|1\rangle$, which is always the case for the input states in \eqref{Wop12}). For $k=1$, the middle qubits in \eqref{T12action} are omitted.
The corresponding circuit diagrams are shown in Fig. \ref{fig:T12ops},
where here $R(\theta)$ is the $R^y(-\theta)$-gate
\begin{equation}
R(\theta)  
 = \begin{pmatrix}
       \cos(\theta/2) & \sin(\theta/2) \\[0.1 cm]
       -\sin(\theta/2) &  \cos(\theta/2)
\end{pmatrix} \,, 
\label{ugate}
\end{equation}
and the angle $\theta_1$ is chosen such that
\begin{equation}
    \cos(\theta_1/2) = c^{(1/2)}_{m,k,1} \,.
\end{equation}

Throughout the paper, we label $m$-qudit vector spaces from $0$ to $m-1$, going from right to left, as in \eqref{bbS}; and in circuit diagrams, the $m$ vector spaces are represented by corresponding wires labeled from the top $(0)$ to the bottom $(m-1)$.

\subsection{Simplifying the circuit}\label{sec:simple12}

We have seen that the operator $U^{(1/2)}_n$ is given by \eqref{Uresult}, \eqref{Wresult} and \eqref{T12action}. According to \eqref{Uop}, 
this operator generates the Dicke states $|D_{n,k}\rangle$ for \emph{all} possible values of $k$. However, if we are only interested in a Dicke state for a \emph{fixed} value of $k$, then it can be shown that some of the $T$ operators become redundant; by removing those redundant operators, we are left with a ``simplified'' $k$-dependent operator $\mathcal{U}^{(1/2)}_{n,k}$
that creates the desired state
\begin{equation}
    |D_{n,k}\rangle = \mathcal{U}^{(1/2)}_{n,k}\, 
     |\psi^{(1/2)}_{n,k}\rangle \,.
\end{equation}
This simplified operator is expressed similarly to \eqref{Uresult}
in terms of corresponding simplified operators $\mathcal{W}^{(1/2)}_{m,k}$ \begin{equation}
\mathcal{U}^{(1/2)}_{n,k} =  \overset{\curvearrowright}{\prod_{m=2}^{n}} 
\left(\mathcal{W}^{(1/2)}_{m,k} \otimes  \id^{\otimes(n-m)} \right)\,,
\label{Uresultsimp12}
\end{equation}
where \cite{Nepomechie:2023lge, Raveh:2023iyy}
\begin{equation}
   \mathcal{W}^{(1/2)}_{m,k} = 
   \overset{\curvearrowleft}{\prod\limits_{k'={\rm max}(k+m-n,1)}^{{\rm min}(k,m-1)}}
\,T^{(1/2)}_{m,k'} \,,
\label{Wresultsimp12}
\end{equation}
cf. \eqref{Wresult}. 

\section{The case $s=1$}\label{sec:spin1}

We turn now to the construction of the spin-1 $T$ operators.
For $s=1$, we see from \eqref{kli1} that $i$ can have two possible values:
either $i=0$ ($k$ is even and $\ell=k/2$), or $i=1$ ($k$ is odd and $\ell=(k-1)/2$).
Correspondingly, there are two families of reference states 
\begin{align}
    |\psi^{(1)\, 0}_{m; \ell}\rangle &= |0\rangle^{\otimes(m-\ell)} |2\rangle^{\otimes \ell} \,, \nonumber \\
    |\psi^{(1)\, 1}_{m; \ell}\rangle &= |0\rangle^{\otimes(m-\ell-1)}|1\rangle |2\rangle^{\otimes \ell} \,,
    \label{refstate1}
\end{align}
where the subscripted semicolon notation is defined in \eqref{refstate} .
The action of the $W$ operator on these reference states is given, according to \eqref{Wop}, by
\begin{align}
    W^{(1)}_m\, |0\rangle^{\otimes(m-\ell)} |2\rangle^{\otimes \ell}
    &=  c^{(1)}_{m,k,0}\, 
    |0\rangle^{\otimes(m-\ell-1)} |2\rangle^{\otimes \ell}|0\rangle
    + c^{(1)}_{m,k,1}\,  |0\rangle^{\otimes(m-\ell-1)}|1\rangle |2\rangle^{\otimes (\ell-1)}|1\rangle  \nonumber \\
    &\qquad + c^{(1)}_{m,k,2}\,  |0\rangle^{\otimes(m-\ell)}|2\rangle^{\otimes \ell}\,,  \qquad k=2\ell \,,
    \label{Wop1even}\\
     W^{(1)}_m\, |0\rangle^{\otimes(m-\ell-1)}|1\rangle 
     |2\rangle^{\otimes \ell}
    &=  c^{(1)}_{m,k,0}\, 
    |0\rangle^{\otimes(m-\ell-2)} |1\rangle |2\rangle^{\otimes \ell}|0\rangle
    + c^{(1)}_{m,k,1}\,  |0\rangle^{\otimes(m-\ell-1)} |2\rangle^{\otimes \ell}|1\rangle  \nonumber \\
    &\qquad + c^{(1)}_{m,k,2}\,  |0\rangle^{\otimes(m-\ell-1)}|1\rangle |2\rangle^{\otimes \ell}\,, \qquad k=2\ell+1\,,
    \label{Wop1odd}
\end{align}
respectively. We will treat these two cases separately in turn, see Eqs. \eqref{Top1even} and \eqref{Top1odd} below.

In order to implement these operators, we make use of the ternary quantum logic gates defined in \cite{Di:2011} (see also \cite{Wang:2020}).
In particular, we denote by $X^{(i,j)}$, with $i<j$, the NOT 
gate that performs the interchange $|i\rangle \leftrightarrow 
|j\rangle$ and leaves unchanged the remaining basis vector.
Similarly, $R^{(i,j)}(\theta)$ denotes the gate that 
performs an $R^{y}(-\theta)$ rotation in the subspace spanned by $|i\rangle$ and $|j\rangle$; hence,
\begin{align}
R^{(i,j)}(\theta) |i \rangle &= \cos(\theta/2) |i \rangle - 
\sin(\theta/2) |j \rangle \,, \non \\
R^{(i, j)}(\theta) |j \rangle &= \sin(\theta/2) |i \rangle + 
\cos(\theta/2) |j \rangle \,.
\label{Rgates}
\end{align}
Moreover, $\begin{quantikz}\gateO{i}\end{quantikz}$ denotes 
a control (of a controlled gate) with value $i$.

\begin{figure}[bht]
	\centering
\begin{adjustbox}{width=0.8\textwidth}
\begin{quantikz}
\lstick{$0$} & \gateO{2} \vqw{3} & \gate{R^{(1,2)}(\theta_1)} \vqw{2} 
& \gateO{2} \vqw{3}  & \gateO{1} \vqw{3} 
& \gate{R^{(0,1)}(\theta_2)} \vqw{2} & \gateO{1} \vqw{3}  & \qw \\
\vdots \\
\lstick{$\ell-1$} & \qw  & \gateO{2} \vqw{1}  & \qw & \qw 
& \gateO{2} \vqw{1} & \qw  & \qw  \\
\lstick{$\ell$} & \gate{X^{(0,1)}}  &  \gateO{1}  
& \gate{X^{(0,1)}} & \gate{X^{(1,2)}}  & \gateO{2} 
& \gate{X^{(1,2)}} & \qw \\
\vdots\\
\lstick{$m-1$}&\qw  &\qw  &\qw  &\qw  &\qw &\qw  & \qw  
\end{quantikz}
\end{adjustbox}
\caption{Circuit diagram for $T^{(1)}_{m,k}= T^{(1)}_{m,2\ell} = T^{(1)\, 0}_{m; \ell}$ ($k$ even), with $1<\ell \le m-1, m>2$}
\label{fig:T1even}
\end{figure}

For $k$ even, we define the 3-qutrit operator 
$T^{(1)}_{m,k}= T^{(1)}_{m,2\ell} \equiv T^{(1)\, 0}_{m\,; \ell}$ 
that performs the mapping \eqref{prop2} with \eqref{Wop1even} as follows
\begin{align}
    T^{(1)\, 0}_{m; \ell}:\quad |0\rangle_\ell\, |2\rangle_{\ell-1}\, 
|2\rangle_0 & \mapsto 
c^{(1)}_{m,k,0}\, |2\rangle_\ell\, |2\rangle_{\ell-1}\, 
|0\rangle_0 +  c^{(1)}_{m,k,1}\, |1\rangle_\ell\, |2\rangle_{\ell-1}\, 
|1\rangle_0 \nonumber \\
&+ c^{(1)}_{m,k,2}\, |0\rangle_\ell\, |2\rangle_{\ell-1}\, 
|2\rangle_0 \,,  \qquad k=2\ell \,,
\label{Top1even}
\end{align}
and otherwise acts as identity (as long as the $0$th qutrit is in the state $|2\rangle$, which is always the case for the input states in \eqref{Wop1even}). For $\ell=1$, the middle qutrits in \eqref{Top1even} are omitted. The corresponding circuit diagram (with $1<\ell \le m-1$ and $m>2$)
is given in Fig. \ref{fig:T1even}, where the rotation angles $\theta_1$ and $\theta_2$ are chosen such that
\begin{align}
\cos(\theta_{1}/2) &= c^{(1)}_{m,k,2}\,, \qquad
\sin(\theta_{1}/2)\cos(\theta_{2}/2) =  c^{(1)}_{m,k,1}
\,, \non \\
& \sin(\theta_{1}/2)\sin(\theta_{2}/2) = c^{(1)}_{m,k,0} \,.
\label{theta1theta2}
\end{align}

The circuit for the edge cases with $l=1$ and $m>1$ can be obtained as a limit of the circuit in Fig. \ref{fig:T1even}, see
Appendix \ref{sec:edge}. 

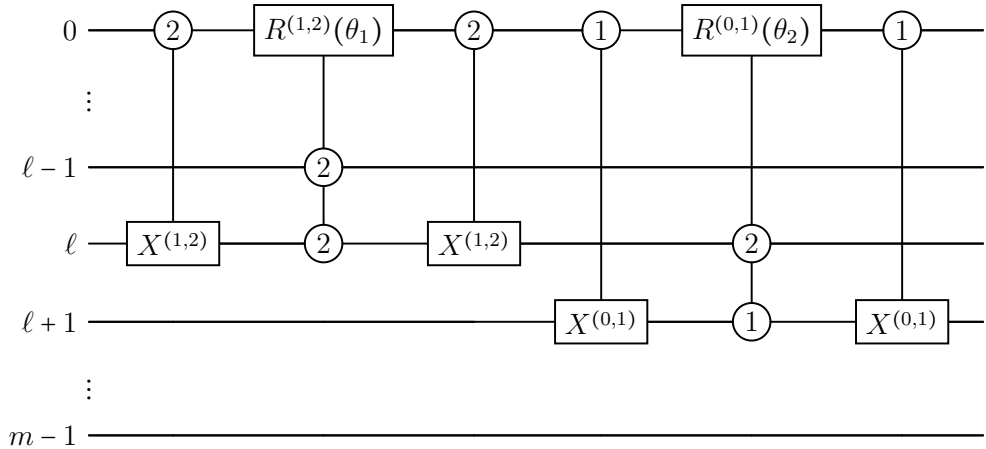
\begin{figure}[htb]
	\centering
\begin{adjustbox}{width=0.8\textwidth}
\begin{quantikz}
\lstick{$0$} & \gateO{2} \vqw{3} & \gate{R^{(1,2)}(\theta_1)} \vqw{2} 
& \gateO{2} \vqw{3}   & \gateO{1} \vqw{4} 
& \gate{R^{(0,1)}(\theta_2)} \vqw{3} & \gateO{1} \vqw{4}  & \qw \\
\vdots \\
\lstick{$\ell-1$} & \qw  & \gateO{2} \vqw{1}  & \qw & \qw & \qw 
& \qw  & \qw  \\
\lstick{$\ell$} & \gate{X^{(1,2)}}  &  \gateO{2}  
& \gate{X^{(1,2)}} & \qw  & \gateO{2} \vqw{1}
& \qw & \qw \\
\lstick{$\ell+1$} & \qw  &\qw  &\qw  & \gate{X^{(0,1)}} &\gateO{1} 
& \gate{X^{(0,1)}}  & \qw \\
\vdots\\
\lstick{$m-1$}&\qw  &\qw  &\qw  &\qw &\qw &\qw &\qw 
\end{quantikz}
\end{adjustbox}
\caption{Circuit diagram for $T^{(1)}_{m,k}= T^{(1)}_{m,2\ell+1} = T^{(1)\, 1}_{m; \ell}$ ($k$ odd), with $1<\ell < m-1, m>3$}
\label{fig:T1odda}
\end{figure}

For $k$ odd, we similarly define a 4-qutrit operator 
$T^{(1)}_{m,k}= T^{(1)}_{m,2\ell+1} \equiv T^{(1)\, 1}_{m\,; \ell}$ 
that performs the mapping \eqref{prop2} with \eqref{Wop1odd} as follows
\begin{align}
    T^{(1)\, 1}_{m; \ell}:\quad |0\rangle_{\ell+1}\, |1\rangle_\ell\, 
|2\rangle_{\ell-1}\, |2\rangle_0 & \mapsto 
c^{(1)}_{m,k,0}\, |1\rangle_{\ell+1}\, |2\rangle_\ell\, 
|2\rangle_{\ell-1}\, |0\rangle_0 
+  c^{(1)}_{m,k,1}\, |0\rangle_{\ell+1}\, |2\rangle_\ell\, 
|2\rangle_{\ell-1}\, |1\rangle_0 \nonumber \\
&+ c^{(1)}_{m,k,2}\, |0\rangle_{\ell+1}\, |1\rangle_\ell\, 
|2\rangle_{\ell-1}\, |2\rangle_0 \,,   \qquad k=2\ell+1 \,.
\label{Top1odd}
\end{align}
The corresponding circuit diagram (with $1<\ell < m-1$ and $m>3$)
is given in Fig. \ref{fig:T1odda}, where the rotation angles $\theta_1$ and $\theta_2$ are again given by \eqref{theta1theta2}.

For the four types of edge cases for $T^{(1)\, 1}_{m; \ell}$ 
\begin{enumerate}[label=(\roman*)]
    \item $\ell=m-1$\,, $m>2$
    \item $\ell=1$\,, $m=2$
    \item $\ell=1$\,, $m>3$
    \item $\ell=0$\,, $m>1$
\end{enumerate}
the corresponding circuit diagrams 
can be obtained from limits of Fig. \ref{fig:T1odda},
see Appendix \ref{sec:edge}.

One can check that the $T$ operators defined by these circuits indeed also satisfy the property \eqref{prop3}. Code in cirq \cite{cirq} for simulating these circuits is included in the Supplementary Material. 

\subsection{Simplifying the circuit}\label{sec:simple1}

As discussed for the case $s=1/2$ in Sec. \ref{sec:simple12}, for a \emph{fixed} value of $k$, not all $T$ operators are needed; by removing the redundant operators, we are left with a ``simplified'' $k$-dependent operator $\mathcal{U}^{(1)}_{n,k}$
that generates the desired state
\begin{equation}
    |D^{(1)}_{n,k}\rangle = \mathcal{U}^{(1)}_{n,k}\, 
     |\psi^{(1)}_{n,k}\rangle \,,
\end{equation}
where
\begin{equation}
\mathcal{U}^{(1)}_{n,k} =  \overset{\curvearrowright}{\prod_{m=2}^{n}} 
\left(\mathcal{W}^{(1)}_{m,k} \otimes  \id^{\otimes(n-m)} \right)\,,
\label{Uresultsimp1}
\end{equation}
and
\begin{equation}
   \mathcal{W}^{(1)}_{m,k} = 
   \overset{\curvearrowleft}{\prod\limits_{k'={\rm max}(k+2(m-n),1)}^{{\rm min}(k,2m-1)}}
\,T^{(1)}_{m,k'} \,.
\label{Wresultsimp1}
\end{equation}

\section{The spin-$s$ case}\label{sec:spins}

We now turn to the construction of the $T$ operators for general values of spin. For spin $s$, there are $2s$ possible values of $i$ in \eqref{kli1}.
We observe from \eqref{Wop} and \eqref{prop2} that
\begin{equation}
   T^{(s)\, i}_{m; \ell}\, |\psi^{(s)\, i}_{m; \ell}\rangle
   = \sum_{j=0}^{2s} c^{(s)}_{m,2s\ell + i,j}\, |\psi^{(s)}_{m-1,2s\ell + i-j}\rangle \otimes |j\rangle \,.
   \label{spinsstep}
\end{equation}
Recalling that the reference states are given by \eqref{refstate}, and noting that 
\begin{equation}
    |\psi^{(s)}_{m-1,2s\ell + i-j}\rangle = \begin{cases}
        |\psi^{(s)\, i-j}_{m-1;\ell}\rangle & i \ge j \\[0.1 in]
        |\psi^{(s)\, 2s+i-j}_{m-1;\ell-1}\rangle & i < j
    \end{cases} \,,
\end{equation}
we see that \eqref{spinsstep} becomes
\begin{align}
   T^{(s)\, i}_{m; \ell}\, |0\rangle^{\otimes(m-\ell-1)}|i\rangle |2s\rangle^{\otimes \ell}
   &= \sum_{j=0}^{i} c^{(s)}_{m,2s\ell + i,j}\,
   |0\rangle^{\otimes(m-\ell-2)}|i-j\rangle |2s\rangle^{\otimes \ell} |j\rangle \nonumber \\
   &+\sum_{j=i+1}^{2s} c^{(s)}_{m,2s\ell + i,j}\,
   |0\rangle^{\otimes(m-\ell-1)}|2s+i-j\rangle |2s\rangle^{\otimes (\ell-1)} |j\rangle  \,.
   \label{Ts}
\end{align}

\begin{figure}[htb]
	\centering
\begin{adjustbox}{width=1.0\textwidth}
\begin{quantikz}
\lstick{$0$}  & \gateO{i} \vqw{4}  
& \gate{R^{(i-1,i)}(\theta_{2s+1-i})} \vqw{3} & \gateO{i} \vqw{4}
& \gateO{i-1} \vqw{4}  
& \gate{R^{(i-2,i-1)}(\theta_{2s+2-i})} \vqw{3} & \gateO{i-1} \vqw{4} 
& \qw & \cdots
& \gateO{1} \vqw{4}  & \gate{R^{(0,1)}(\theta_{2s})} \vqw{3} 
& \gateO{1} \vqw{4} & \qw \\
\vdots \\
\lstick{$\ell-1$}  & \qw & \qw & \qw
& \qw & \qw & \qw & \qw & \cdots 
& \qw & \qw & \qw & \qw \\
\lstick{$\ell$}   & \qw & \gateO{2s} \vqw{1} & \qw
& \qw & \gateO{2s} \vqw{1} & \qw & \qw & \cdots 
& \qw & \gateO{2s} \vqw{1} & \qw & \qw \\
\lstick{$\ell+1$}  & \gate{X^{(0,1)}} & \gateO{1} & \gate{X^{(0,1)}} 
& \gate{X^{(1,2)}} & \gateO{2} & \gate{X^{(1,2)}} 
& \qw & \cdots 
& \gate{X^{(i-1,i)}} & \gateO{i} & \gate{X^{(i-1,i)}} 
& \qw \\
\vdots\\
\lstick{$m-1$}  &  \qw & \qw & \qw
& \qw & \qw & \qw & \qw & \cdots
& \qw & \qw & \qw & \qw
\end{quantikz}
\end{adjustbox}
\caption{Circuit diagram for $T^{(s)}_{m,k} = T^{(s)}_{m,2s\ell + i} = T^{(s)\, i}_{m; \ell}$ (end). Note that this part of the circuit is present if and only if $i>0$.}
\label{fig:Tsbis}
\end{figure}
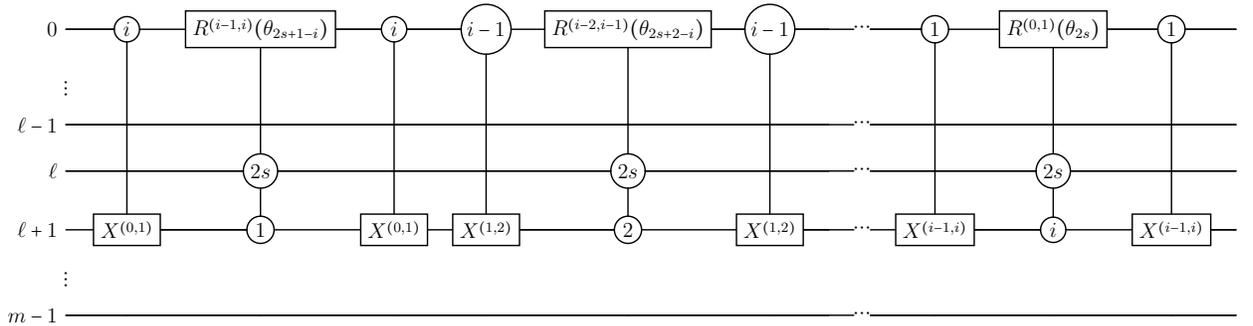

Our circuit for $T^{(s)\, i}_{m; \ell}$ generates the terms in \eqref{Ts} from last to first; that is, starting from the reference state on the l.h.s, the circuit successively generates the terms on the r.h.s with $j=2s\,, j=2s-1\,, \ldots, j=0$. The circuit diagram, split into two parts due to its length, is shown  Fig. \ref{fig:Ts} (beginning) and Fig. \ref{fig:Tsbis} (end). If $i=0$, then the circuit ends at the dashed red line in Fig.  \ref{fig:Ts}; otherwise, the circuit continues through Fig. \ref{fig:Tsbis}. The $2s$ rotation angles $\theta_1, \ldots, \theta_{2s}$ are chosen such that
\begin{equation}
    \sin(\theta_1/2) \, \cdots \sin(\theta_{2s-j}/2)\,  \cos(\theta_{2s+1-j}/2) =  c^{(s)}_{m,2s\ell + i,j} \,, \qquad j = 0, 1, \ldots, 2s\,,
\end{equation}
where $\theta_{2s+1} \equiv 0$. The displayed circuit diagram is for the generic case with $1 < \ell < m-1$ and $m>3$; edge cases can be obtained from limits, as for $s=1/2$ and $s=1$.

One can easily check that, for $s=1/2$ and $s=1$,
this circuit reduces to the ones presented in Secs. \ref{sec:spin12} and \ref{sec:spin1}, respectively. 
Code in cirq \cite{cirq} for simulating the circuit for $s=3/2$
is also included in the Supplementary Material.

We observe from Figs. \ref{fig:Ts}-\ref{fig:Tsbis} that $T^{(s)\, i}_{m; \ell}$ is generically a 4-qudit operator (3-qudit for $i=0$); i.e., the number of qudits on which it acts does not grow with $s$. Moreover,  $T^{(s)\, i}_{m; \ell}$ generically has $2s$ double-controlled rotations, and $4s$ single-controlled NOTs. Recalling that the number of $T$ operators
in $U^{(s)}_n$ is $\mathcal{O}(n^2 s)$ \eqref{Tcount}, we conclude that the total number of gates in $U^{(s)}_n$ is $\mathcal{O}(n^2 s^2)$.

We note that the double-controlled rotation gates can be decomposed
into elementary 1-qudit and 2-qudit gates in the same way as for
corresponding double-controlled {\it qubit} $R^y$ gates, since we use
the naive embeddings $SU(2) \subset SU(2s+1)$ \eqref{Rgates}.  Hence, eight
2-qudit gates are needed for the decomposition of each double-controlled rotation gate, see e.g. Table 1 in \cite{Nepomechie:2023lge}. Universal gate sets for qudit-based quantum computing are reviewed in \cite{Wang:2020}.

\begin{figure}[htb]
	\centering
\begin{adjustbox}{width=1.0\textwidth}
\begin{quantikz}
\lstick{$0$} & \gateO{2s} \vqw{3} & \gate{R^{(2s-1,2s)}(\theta_1)} \vqw{2} 
& \gateO{2s} \vqw{3}   & \gateO{2s-1} \vqw{3} 
& \gate{R^{(2s-2,2s-1)}(\theta_2)} \vqw{2} & \gateO{2s-1} \vqw{3} 
& \cdots 
& \gateO{i+1} \vqw{3} & \gate{R^{(i,i+1)}(\theta_{2s-i})} \vqw{2} 
& \gateO{i+1} \vqw{3} \slice{stop if $i=0$}  & \qw  \\
\vdots \\
\lstick{$\ell-1$} & \qw  & \gateO{2s} \vqw{1}  & \qw & \qw & \gateO{2s} \vqw{1} 
& \qw  & \cdots
& \qw  & \gateO{2s} \vqw{1}  & \qw & \qw  \\
\lstick{$\ell$} & \gate{X^{(i,i+1)}}  &  \gateO{i+1}  
& \gate{X^{(i,i+1)}} & \gate{X^{(i+1,i+2)}}  & \gateO{i+2} 
& \gate{X^{(i+1,i+2)}}  & \cdots 
& \gate{X^{(2s-1,2s)}}  & \gateO{2s}  
& \gate{X^{(2s-1,2s)}}  & \qw  \\
\lstick{$\ell+1$} & \qw  &\qw  &\qw  & \qw & \qw
& \qw   & \cdots 
& \qw  &\qw  &\qw & \qw \\
\vdots\\
\lstick{$m-1$}&\qw  &\qw  &\qw  &\qw &\qw &\qw  & \cdots
& \qw  &\qw  &\qw & \qw  
\end{quantikz}
\end{adjustbox}
\caption{Circuit diagram for $T^{(s)}_{m,k} = T^{(s)}_{m,2s\ell + i} = T^{(s)\, i}_{m; \ell}$ (beginning)}
\label{fig:Ts}
\end{figure}
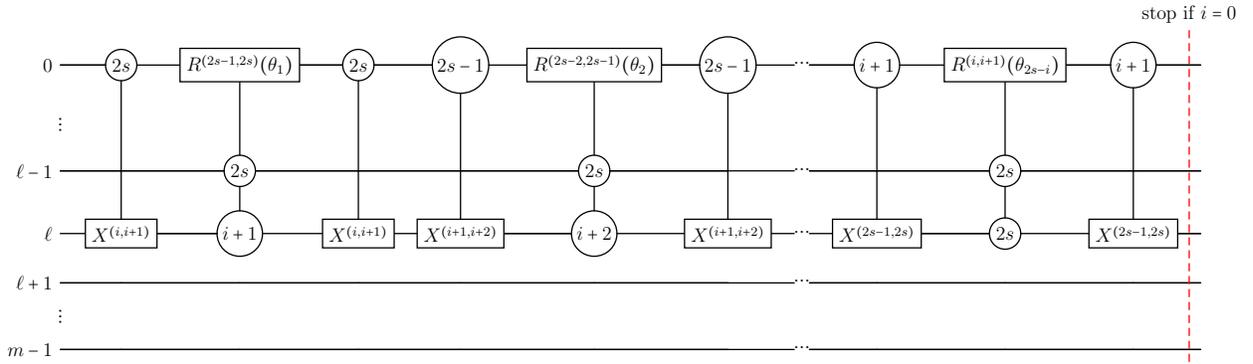

\subsection{Simplifying the circuit}\label{sec:simples}

As discussed for the cases $s=1/2$ and $s=1$
in Secs. \ref{sec:simple12} and \ref{sec:simple1} respectively,
for a \emph{fixed} value of $k$, not all $T$ operators are needed; by removing the redundant operators, we are left with a ``simplified'' $k$-dependent operator $\mathcal{U}^{(s)}_{n,k}$
that generates the desired state
\begin{equation}
    |D^{(s)}_{n,k}\rangle = \mathcal{U}^{(s)}_{n,k}\, 
     |\psi^{(s)}_{n,k}\rangle \,,
     \label{simplifieds}
\end{equation}
where
\begin{equation}
\mathcal{U}^{(s)}_{n,k} =  \overset{\curvearrowright}{\prod_{m=2}^{n}} 
\left(\mathcal{W}^{(s)}_{m,k} \otimes  \id^{\otimes(n-m)} \right)\,,
\label{Uresultsimps}
\end{equation}
and
\begin{equation}
   \mathcal{W}^{(s)}_{m,k} = 
   \overset{\curvearrowleft}{\prod\limits_{k'={\rm max}(k+2s(m-n),1)}^{{\rm min}(k, 2sm-1)}}
\,T^{(s)}_{m,k'} \,,
\label{WWs}
\end{equation}
see \eqref{Wresultsimp12}, \eqref{Wresultsimp1}, where again the $T$ operators are given by Figs. \ref{fig:Ts}, \ref{fig:Tsbis}.

The number of $T$ operators in $\mathcal{U}^{(s)}_{n,k}$ is given, in view of \eqref{Uresultsimps} and \eqref{WWs}, by
\begin{align}
  N^{(s)}_{n,k} &= \sum_{m=2}^n \left[1 +
    {\rm min}(k, 2 s m-1) - {\rm max}(k+ 2s(m-n),1) \right]\non \\
    &= \mathcal{O}(k(2sn-k))
    \,,
    \label{Tcounts}
\end{align}
which can be shown to be consistent with the duality symmetry \eqref{duality}
\begin{equation}
  N^{(s)}_{n,k} = N^{(s)}_{n, 2sn - k} \,.  
\end{equation}
Hence,
the total number of gates in $\mathcal{U}^{(s)}_n$ is 
$\mathcal{O}(sk(2sn-k))$.

\section{Discussion}\label{sec:discuss}

We have introduced the notion of $su(2)$ spin-$s$ Dicke states $|D^{(s)}_{n,k}\rangle$ \eqref{spinsDicke}, \eqref{closed}, which are higher-spin generalizations of usual (spin-1/2) Dicke states that can be decomposed into linear combinations of $su(2s+1)$ qudit Dicke states. Based on the recursive property \eqref{recursion}, we have formulated a circuit for preparing these states. (Specifically, 
we have determined a $k$-independent operator $U^{(s)}_n$ that generates via \eqref{Uop} these spin-$s$ Dicke states from reference states $|\psi^{(s)}_{n,k}\rangle$ in terms of $W$'s \eqref{Uresult}, which are in turn given in terms of $T$'s \eqref{Wresult}; and the $T$'s are (at most) 4-qudit operators given by the circuits in Figures \ref{fig:Ts} and \ref{fig:Tsbis}.) These Dicke states can also be generated with a ``simplified'' $k$-dependent operator $\mathcal{U}^{(s)}_{n,k}$ using fewer $T$ operators, see  \eqref{simplifieds}-\eqref{WWs}. We emphasize that this algorithm 
for preparing Dicke states $|D^{(s)}_{n,k}\rangle$
is deterministic, does not use ancillary qudits, and the number of gates scales as $sk(2sn-k)$, see \eqref{Tcounts}; this circuit is therefore efficient \cite{Nielsen:2019} to the extent that its size is polynomial in the system size $n$ (as well as the spin $s$ and the parameter $k$).
These circuits are significantly simpler than those for 
preparing $(2s+1)$-level qudit Dicke states \cite{Nepomechie:2023lge}.

We have also precisely related spin-$s$ Dicke states to $(2s+1)$-level qudit Dicke states \eqref{decomp},  \eqref{alphas}, and we have computed their entanglement entropy \eqref{EE}, \eqref{EEapprox}. Because spin-$s$ Dicke states are linear combinations of qudit Dicke states, one could use quantum phase estimation to project a spin-$s$ Dicke state onto a desired qudit Dicke state, with success probability $|\alpha_{n,k}^{(s)}(\vec k)|^2$ \eqref{alphas}, see  \cite{Wang:2021, Piroli:2024ckr}. 

Further properties and applications of spin-$s$ Dicke states remain 
to be explored. For example, it would be interesting to consider 
their $q$-deformation, and compare with corresponding results for 
$(2s+1)$-level qudit Dicke states. Indeed, it was shown 
\cite{Raveh:2023iyy} that celebrated $q$-combinatorial identities arise naturally from the $q$-analog qudit Dicke states; perhaps other identities can be related to the $q$-analog of the spin-$s$ Dicke 
states, such as (possibly) a $q$-analog of \eqref{combidentity}. 
Moreover, as noted in the Introduction, these states may be 
useful for generalizing the many applications of (qubit) Dicke states 
to qudits, and may serve as the starting point for preparing exact 
eigenstates of integrable higher-spin Heisenberg chains.

\section*{Acknowledgments} 
We thank Matthias Christandl for discussions. 
RN acknowledges the hospitality and financial support
from INFN BO and Bologna University (ISA Visiting Fellowship), where this work was initiated. RN is also supported in part by the National Science Foundation under grant PHY 2310594, and by a Cooper fellowship. FR thanks the RFO funds of the Dept. of Physics and Astronomy of Bologna University and INFN for financial support through the GAST and DOT4 grants. 

\appendix

\section{Spin-$s$ Dicke states in terms of $(2s+1)$-level qudit Dicke states}\label{sec:relating}

We derive here the closed-form expression \eqref{closed} for the spin$s$ Dicke states by expressing the $su(2)$ spin-$s$ Dicke states in terms of $su(2s+1)$ qudit Dicke states, see \eqref{decomp} below.

\subsection{Qudit Dicke states}

We begin by defining a multiset $M(\vec k)$
\begin{equation}
M(\vec k)	=\{ \underbrace{0, \ldots, 0}_{k_{0}}, \underbrace{1, 
\ldots, 1}_{k_{1}}, \ldots, \underbrace{d-1, \ldots, d-1}_{k_{d-1}}\} 
\,,
\label{multiset}
\end{equation}
where $k_{j}$ is the multiplicity of $j$ in $M(\vec k)$, such that $M(\vec 
k)$ has cardinality $n$. Hence, $\vec k$ is a $d$-dimensional vector such that
\begin{equation}
\vec k = (k_{0}, k_{1}, \ldots, k_{d-1})\quad \text{with}\quad k_{j} \in \{0, 
1, \ldots, n\}\quad \text{and}\quad \sum_{j=0}^{d-1} k_{j} = n\,.
\label{ks}
\end{equation}

The corresponding normalized qudit Dicke state $|D^{n}(\vec k)\rangle$ with
a number $n$ of $d$-level qudits is defined by (see \cite{Nepomechie:2023lge} and references therein)
\begin{equation}
|D^{n}(\vec k)\rangle  = \frac{1}{\sqrt{\binom{n}{\vec k}}}
\sum_{w \in \mathfrak{S}_{M(\vec k)}}  | w \rangle \,,
\label{Dickedef}
\end{equation}
where $\mathfrak{S}_{M(\vec k)}$ is the set of permutations of the 
multiset $M(\vec k)$ \eqref{multiset},  
and $| w  \rangle$ is the $n$-qudit state corresponding 
to the permutation $w$. Moreover, $\binom{n}{\vec k}$ denotes the multinomial
\begin{equation}
\binom{n}{\vec k} = \binom{n}{k_{0}, k_{1}, \ldots, k_{d-1}}	= 
\frac{n!}{\prod_{j=0}^{d-1}k_{j}!} \,.
\label{size}
\end{equation}
The qudit Dicke states satisfy the recursion \cite{Nepomechie:2023lge}
\begin{equation}
|D^{n}(\vec k)\rangle  = \sum_{j=0}^{d-1}
\sqrt{\frac{k_j}{n}}
| D^{n-1}(\vec k-\hat j) \rangle\otimes|j\rangle \,,
\label{quditrec}
\end{equation}
where $\hat j$ is the $d$-dimensional unit vector with components $(\hat j)_a=\delta_{aj},\,a=0,1,\ldots,d-1$.

\subsection{Higher-spin Dicke states in terms of qudit Dicke states}

As a simple example of a spin-$s$ Dicke state in terms of $(2s+1)$-level qudit Dicke states, we observe that the spin-1 Dicke state \eqref{example} can be rewritten as
\begin{equation}
   |D^{(1)}_{3, 2}\rangle = \frac{2}{\sqrt{5}} |D^{3}(1,2,0)\rangle +  \frac{1}{\sqrt{5}} |D^{3}(2,0,1)\rangle \,.
   \label{example2}
\end{equation}
We now proceed to generalize this result.

Because the spin-$s$ Dicke state $|D^{(s)}_{n,k}\rangle$  is invariant under any permutation, we know it can be decomposed into a linear combination of the permutation-invariant $d$-level qudit Dicke states $|D^{n}(\vec k)\rangle$, as in \eqref{example2}. Evidently, we need $d=2s+1$.
Moreover, we observe that such a qudit Dicke state is an eigenstate of $\mathbb{S}^z$ with eigenvalue $sn-\sum_{j=0}^{2s}j\,k_j$; and since the $\mathbb{S}^z$ eigenvalue of the spin-$s$ Dicke state $|D^{(s)}_{n,k}\rangle$ is given by $sn-k$, the decomposition is restricted to $\vec k$'s that satisfy
\begin{equation}
    \sum_{j=0}^{2s} j\, k_j  = k  \,.
    \label{D1}
\end{equation}
We also require that the number of qudits match, so
\begin{equation}
    \sum_{j=0}^{2s} k_{j} = n \,, 
    \label{D2}
\end{equation}
see \eqref{ks}. 
In other words, for given values of $n$, $k$ and $s$,
the allowed values of $\vec k = (k_{0}, k_{1}, \ldots, k_{2s})$ are precisely the solutions of the Diophantine equations \eqref{D1} and \eqref{D2}. The number of such solutions, which we denote by $g^{(s)}_{n,k}$, is known to be generated by the $q$-binomial coefficient \footnote{Indeed, in the language of \cite{Stanley2011} on page 65, the multiset $M(\vec k)$ \eqref{multiset} with $d=2s+1$ defines a \emph{partition} of $k$ (defined in Eq. \eqref{D1}), and the nonzero 
elements of $M(\vec k)$ are its \emph{parts}; according to Proposition 1.7.3 in \cite{Stanley2011}, the number of partitions of $k$ into at most $n$ parts, with largest part at most $2s$, is given by $g^{(s)}_{n,k}$ in \eqref{gs}. We thank Michelle Wachs for pointing us to this reference.}
\begin{equation}
   \binom{n+2s}{2s}_q =  \sum_{k=0}^{2  s n} g^{(s)}_{n,k}\, q^k \,.
   \label{gs}
\end{equation}
We conclude that a spin-$s$ Dicke state has the decomposition
\begin{equation}
   |D^{(s)}_{n,k}\rangle  = \sideset{}{'}\sum_{\vec k} \alpha^{(s)}_{n,k}(\vec k)\, |D^{n}(\vec k)\rangle \,,
   \label{decomp}
\end{equation}
where the sum over $\vec k$ is restricted (indicated by a prime) to solutions of \eqref{D1} and \eqref{D2}, and the coefficients $\alpha^{(s)}_{n,k}(\vec k)$ are still to be determined. 

In order to determine the coefficients $\alpha^{(s)}_{n,k}(\vec k)$, we  substitute \eqref{decomp} into the recursion \eqref{recursion}, and obtain
\begin{equation}
\sideset{}{'}\sum_{\vec k}
\alpha^{(s)}_{n,k}(\vec k)\, |D^{n}(\vec k)\rangle=
\sum_{j=0}^{2s} c_{n,k,j}^{(s)}\,
\sideset{}{''}\sum_{\vec a} \alpha^{(s)}_{n-1,k-j}(\vec a)\, |D^{n-1}(\vec a)\rangle\otimes|j\rangle\,,
\label{D3}
\end{equation}
where on the r.h.s we restrict (indicated by a double-prime) $\vec a$ to be solutions of \eqref{D1} and \eqref{D2} with $k_j\to a_j\,, n\to n-1,\,k\to k-j$. Expanding $|D^{n}(\vec k)\rangle$ in \eqref{D3} via the qudit Dicke state recursion \eqref{quditrec}, and making the association $\vec a\to\vec k-\hat j$ then gives a recursive relation for the coefficients $\alpha^{(s)}_{n,k}(\vec k)$
\begin{equation}
\sqrt{\frac{k_j}{n}}\, \alpha^{(s)}_{n,k}(\vec k)
=c_{n,k,j}^{(s)}\,\alpha^{(s)}_{n-1,k-j}(\vec k-\hat j),
\end{equation}
which solves to
\begin{equation}
\alpha^{(s)}_{n,k}(\vec k)=
\left[\frac{\binom{n}{\vec k}}{\binom{2sn}{k}}\,\prod_{j=0}^{2s}\,
\binom{2s}{j}^{k_j}\right ]^{1/2}.
\label{alphas}
\end{equation}
We note that the orthonormality of the spin-$s$ and qudit Dicke states thus implies the combinatorial identity\footnote{This identity follows from Fa\`a di Bruno's formula when considering $k$ derivatives of $f(g(x))$ with $f(x)=x^n$ and $g(x)=(1+x)^{2s}$ and evaluating at $x=0$. We thank Math Stack Exchange user `ameg' for pointing out to us this proof.}
\begin{equation}
\binom{2sn}{k}=\sum_{\vec k}
\binom{n}{\vec k}\,\prod_{j=0}^{2s}\,
\binom{2s}{j}^{k_j},
\label{combidentity}
\end{equation}
where we sum over solutions $\vec k$ to the Diophantine equations \eqref{D1} and \eqref{D2}. Finally, \eqref{closed} follows from \eqref{Dickedef}, \eqref{decomp} and \eqref{alphas}.

The paper \cite{Liu:2015} includes a discussion of particular cases of the results \eqref{decomp} and \eqref{alphas}. Specifically, for the three cases $s=1, 3/2, 2$, formulas for the coefficients $\alpha^{(s)}_{n,k}(\vec k)$ are given in \cite{Liu:2015} 
(note that our variables $(n, k, \vec k)$ correspond to the variables $(N, J - M, \vec n)$ in \cite{Liu:2015}, with $J = s N$), 
see there Eqs. (13), (25), (34); and sample values of these coefficients are reported in corresponding tables.  Where there is overlap, our results agree with those in \cite{Liu:2015}, apart from some typos in the latter.

\subsection{An alternative construction of spin-$s$ Dicke states}

Since a spin-$s$ Dicke state can be expressed \eqref{decomp} as a linear combination of $(2s+1)$-level qudit Dicke states, the construction   \cite{Nepomechie:2023lge} of the latter  can 
-- in principle -- be used to obtain an alternative construction of the former. Indeed, a unitary operator $U_n$ is found in \cite{Nepomechie:2023lge} that generates the qudit Dicke state \eqref{Dickedef} from a product  state $|{\rm e}(\vec{k})\rangle$
\begin{equation}
    U_n\, |{\rm e}(\vec{k})\rangle=|D^n(\vec k)\rangle\,, \qquad
    |{\rm e}(\vec k) \rangle = |0\rangle^{\otimes k_{0}}|1\rangle^{\otimes k_{1}}\ldots |d-1\rangle^{\otimes k_{d-1}} \,.
    \label{quditDickeOp}
\end{equation}
It is not difficult to prepare the linear combination of the 
$|{\rm e}(\vec{k})\rangle$'s 
\begin{equation}
    |\alpha^{(s)}_{n,k}(\vec k)\rangle = \sideset{}{'}\sum_{\vec k} \alpha^{(s)}_{n,k}(\vec k)\, |{\rm e}(\vec{k})\rangle \,,
    \label{alphavec}
\end{equation}
where the coefficients $\alpha^{(s)}_{n,k}(\vec k)$ are given by \eqref{alphas}.
It follows from \eqref{decomp}, \eqref{quditDickeOp} and \eqref{alphavec} that the spin-$s$ Dicke state can be constructed by acting with $U_n$ on the above state  
\begin{equation}
|D^{(s)}_{n,k}\rangle  = U_n\,  |\alpha^{(s)}_{n,k}(\vec k)\rangle \,.
\end{equation}
(This is similar to the construction of a linear combination of qubit Dicke states in Theorem 2 of \cite{Bartschi2019}.) However, the quantum circuit for $U_n$ in \cite{Nepomechie:2023lge}
is considerably more complicated than the circuit for $U^{(s)}_n$ \eqref{Uop} in the present work.

\section{Proof of the recursion formula \eqref{recursion}}\label{sec:proof}

In this section, we denote the total spin operators \eqref{bbS} by $\vec{\mathbb{S}}^{(n)}$ in order to indicate the number of spins (qudits). It follows from \eqref{bbS} that these operators satisfy the recursion
\begin{equation}
    \vec{\mathbb{S}}^{(n)} = \vec{\mathbb{S}}^{(n-1)} \otimes \id + \id^{\otimes(n-1)} \otimes \vec{S} \,,
\end{equation}
where $\vec{S} = \vec{\mathbb{S}}^{(1)}$.
Therefore, powers of the total spin-lowering operator are given by
\begin{equation}
    \left(\mathbb{S}^{(n)-}\right)^k   
    = \left( \mathbb{S}^{(n-1)-} \otimes \id + \id^{\otimes(n-1)} \otimes S^- \right)^k  
    = \sum_{j=0}^k \binom{k}{j} \left(\mathbb{S}^{(n-1)-}\right)^{k-j} \otimes \left(S^- \right)^j \,.
\end{equation}
Recalling the definition \eqref{spinsDicke} of spin-$s$ Dicke states, we obtain 
\begin{align}
    |D^{(s)}_{n,k}\rangle &= a^{(s)}_{n,k}  \left(\mathbb{S}^{(n)-} \right)^k |0\rangle^{\otimes n} \nonumber\\
    &= a^{(s)}_{n,k} \left\{ 
    \sum_{j={\rm max}(0,k-2s(n-1))}^{{\rm min}(k, 2s)} \binom{k}{j} \left(\mathbb{S}^{(n-1)-}\right)^{k-j} \otimes \left(S^- \right)^j \right\} 
    \left(|0\rangle^{\otimes (n-1)}\otimes |0\rangle\right) \,,
    \label{step}
\end{align}
where the limits in the sum reflects the fact that  $\left(\mathbb{S}^{(n)-}\right)^j\,|0\rangle^{\otimes n}=0$ for $j>2sn$. Noting also that 
\begin{equation}
    a^{(s)}_{1,j}\left(S^- \right)^j |0\rangle =  |j\rangle \,, \qquad j = 0, 1, ..., 2s \,,
\end{equation}
we conclude from \eqref{step} that
\begin{equation}
    |D^{(s)}_{n,k}\rangle = \sum_{j={\rm max}(0,k-2s(n-1))}^{{\rm min}(k, 2s)} c^{(s)}_{n,k,j}\, |D^{(s)}_{n-1,k-j}\rangle \otimes |j\rangle \,,
    \label{recursionfull}
\end{equation}
where
\begin{equation}
  c^{(s)}_{n,k,j} = \binom{k}{j} \frac{a^{(s)}_{n,k}}{a^{(s)}_{n-1,k-j}\, a^{(s)}_{1,j}} 
  = \sqrt{\frac{\binom{2s}{j}\binom{2sn-2s}{k-j}}{\binom{2sn}{k}}} \,,
  \label{ccoefs2}
\end{equation}
where we used the result \eqref{normalization} to pass to the the final equality. Focusing on the limits in the sum in \eqref{recursionfull}, and recalling the definition of $\ell$ \eqref{kli1}, we
note that $k<2s$ implies $\ell=0$, and $k-2s(n-1)>0$ implies $\ell=n-1$ (or the trivial case $\ell=n$). Because the circuits for $\ell=0,1,n-1$ will be treated separately as edge cases of the circuit for $1<\ell<n-1$, instead of \eqref{recursionfull}
we simply write 
\begin{equation}
    |D^{(s)}_{n,k}\rangle = \sum_{j=0}^{2s} c^{(s)}_{n,k,j}\, |D^{(s)}_{n-1,k-j}\rangle \otimes |j\rangle \,,
    \label{recursion2}
\end{equation}
assuming $1<\ell<n-1$.

\section{Entanglement entropy}\label{sec:EE}

Usual (spin-$1/2$) Dicke states have long been known to be entangled; indeed, the simplest such state $|D_{2,1}\rangle = \left(|0 1 \rangle + |1 0 \rangle\right)/\sqrt{2}$ is a Bell state. The Von Neumann bipartite entanglement entropy of a Dicke state $|D_{n,k}\rangle$ for general values of $n$ and $k$ was computed in \cite{Popkov:2004, Latorre:2004qn}, see also \cite{Lucke:2014mcw,
Moreno:2018, Munizzi:2023ihc}. 
Corresponding results were obtained
for qudit Dicke states $|D^{n}(\vec k)\rangle$ \eqref{Dickedef} in \cite{Popkov:2005, Carrasco:2015sxh}, as well as for their $q$-analogs in \cite{Li:2015, Raveh:2023iyy}.

We calculate here the bipartite entanglement entropy of the spin-$s$ Dicke states $|D^{(s)}_{n,k}\rangle$ \eqref{spinsDicke}. This entails partitioning the $n$ qudits into two parts, of sizes $n-l$ and $l$, where $l<n$ is a positive integer, and calculating the eigenvalues of the reduced density matrix of $|D_{n,k}^{(s)}\rangle$\,, obtained by tracing over the first $n-l$ qudits of the density matrix. One method of computing these eigenvalues is to find the Schmidt decomposition for the spin-$s$ Dicke states, which we claim is given by
\begin{equation}
|D^{(s)}_{n,k}\rangle = \sum_{j={\rm max}(0,k-2s(n-l))}^{{\rm min}(k,2sl)} 
\sqrt{\lambda_j}\, 
|D^{(s)}_{n-l,k-j}\rangle \otimes
|D^{(s)}_{l,j}\rangle \,,
\label{Schmidt}
\end{equation}
where 
\begin{equation}
\lambda_{j}=
\frac{\binom{2sl}{j}\binom{2sn-2sl}{k-j}}{\binom{2sn}{k}}\,.
\end{equation}
Note that the max/min in \eqref{Schmidt} are simply enforcing the requirements $0\leq j\leq 2sl$ and $0\leq k-j\leq 2sn-2sl$. A proof for this decomposition can be obtained by generalizing the argument in Appendix \ref{sec:proof}; indeed, the recursion \eqref{recursionfull} can be viewed as a special case of the Schmidt decomposition \eqref{Schmidt} with $l=1$, as we see that $\sqrt{\lambda_{j}}=c_{n,k,j}^{(s)}$ when $l=1$. Using the orthonormality of the spin-$s$ Dicke states, it follows from the Schmidt decomposition that the eigenvalues of the reduced density matrix are given by $\lambda_j$, and therefore the entanglement entropy of $|D^{(s)}_{n,k}\rangle$ is given by
\begin{equation}
    S_l= \sum_{j={\rm max}(0,k-2s(n-l))}^{{\rm min}(k,2sl)} -\lambda_{j}\,\log_{2s+1}\lambda_{j}\,.
\label{EE}
\end{equation}
We observe that $\lambda_j$, and therefore $S_l$, are invariant under $k\to 2sn-k$, corresponding to the duality symmetry \eqref{duality}. This allows us to focus on the case $k\leq sn$. Note that while $\lambda_{j}\equiv\lambda_j(l)$ is not symmetric under $l\to n-l$, it does satisfy $\lambda_j(l)=\lambda_{k-j}(n-l)$. It follows that $S_l$ is also symmetric under $l\to n-l$, so we restrict our attention to $l\leq n/2$. 

Following \cite{Latorre:2004qn}, for large $n$ and $l$, we can approximate the hypergeometric distribution $\lambda_j$ via the Gaussian distribution
\begin{equation}
    \lambda_j\approx\frac{1}{\sqrt{2\pi\sigma}}\exp\left[-\frac{(j-\bar j)^2}{2\sigma^2}\right]\,,
\end{equation}
with mean $\bar j=kl/n$ and variance
\begin{equation}
\sigma^2=k(2sn-k)l(n-l)/2sn^3\,,
\label{variance}
\end{equation}
the latter of which exhibits the expected symmetries $l\to n-l$ and $k\to2sn-k$.
The entanglement entropy is therefore approximated by
\begin{equation}
S_l\approx -\int_{-\infty}^{\infty}
\lambda_{j}\,\log_{2s+1}\lambda_{j}\, dj=\frac{1}{2}\log_{2s+1}(2\pi e\sigma^2)\,.
\label{EEapprox}
\end{equation}

\begin{figure}[htb]
      \centering
\includegraphics[width=0.6\linewidth]{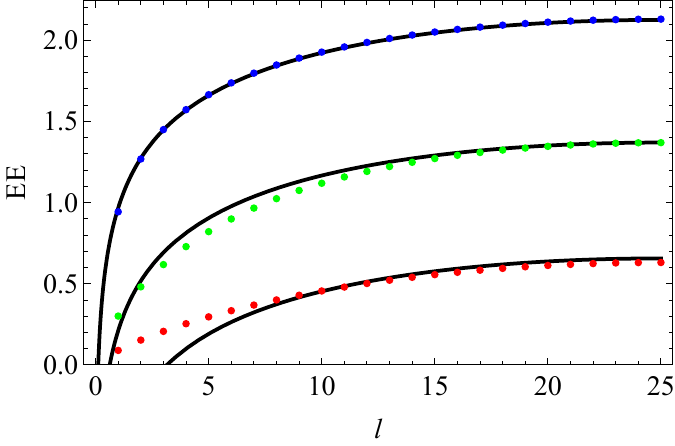}
\caption{The entanglement entropy (EE) of the state $|D^{(s)}_{n,k}\rangle$
as a function of $l$ for $s=1\,, n=50$. The exact values are given for $k=1$ (red), $k=5$ (green), and $k=50$ (blue), as are their respective approximated curves (black).}
\label{fig:EE}
\end{figure}

We plot the entanglement entropy curves---both the numerical sums  given by \eqref{EE} and the approximated curves given by \eqref{EEapprox}---for $s=1$
in Fig. \ref{fig:EE}. The results are qualitatively similar to those for the case $s=1/2$ \cite{Popkov:2004, Latorre:2004qn}. Indeed,
the variance \eqref{variance} for fixed $n,k,s,l$ can be mapped to the variance for the
usual (spin-$1/2$) Dicke state $|D_{2sn,k}\rangle$ with
partitions of sizes $2sl$ and $2sn-2sl$. In other words, 
denoting the result in  \eqref{variance} by $\sigma^2(n,k,s,l)$, we see that
\begin{equation}
\sigma^2(n,k,s,l) = \sigma^2(2sn, k ,\frac{1}{2}, 2sl) \,,
\label{mapping}
\end{equation}
corresponding to the mapping $(n,l) \mapsto (2sn, 2sl)$, i.e. ``stretching'' both the chain and the partition by the factor $2s$.

Comparing the entanglement entropy results for spin-$s$ Dicke states and for  $(2s+1)$-level qudit Dicke states \cite{Popkov:2005, Carrasco:2015sxh}, we find that they cannot be mapped into each other (such as in  \eqref{mapping}) except for $s=1/2$. This is not surprising, given
that the latter states have $2s+1$ free parameters ($\vec k$, where $n=\sum_i k_i$), while the former states have only two ($n$ and $k$).

\clearpage
\section {Circuit diagrams for $s=1$ edge cases}\label{sec:edge}

For completeness, we display here circuit diagrams corresponding to the various edge cases for $s=1$.

\begin{figure}[htb]
	\centering
\begin{adjustbox}{width=0.8\textwidth}
\begin{quantikz}
\lstick{$0$} & \gateO{2} \vqw{1} & \gate{R^{(1,2)}(\theta_1)} \vqw{1} 
& \gateO{2} \vqw{1}  & \gateO{1} \vqw{1} 
& \gate{R^{(0,1)}(\theta_2)} \vqw{1} & \gateO{1} \vqw{1}  & \qw \\
\lstick{$\ell=1$} & \gate{X^{(0,1)}}  &  \gateO{1}  
& \gate{X^{(0,1)}} & \gate{X^{(1,2)}}  & \gateO{2} 
& \gate{X^{(1,2)}} & \qw \\
\vdots\\
\lstick{$m-1$}&\qw  &\qw  &\qw  &\qw  &\qw &\qw  & \qw  
\end{quantikz}
\end{adjustbox}
\caption{Circuit diagram for $T^{(1)}_{m,k}= T^{(1)}_{m,2\ell} = T^{(1)\, 0}_{m; \ell}$ ($k$ even), with $\ell=1, m>1$}
\label{fig:T1evenb}
\end{figure}
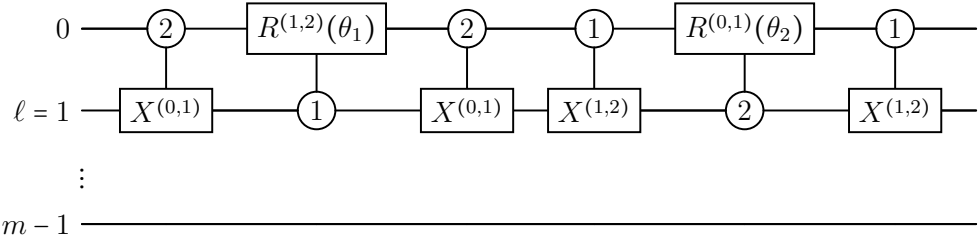

\begin{figure}[htb]
	\centering
\begin{adjustbox}{width=0.5\textwidth}
\begin{quantikz}
\lstick{$0$} & \gateO{2} \vqw{3} & \gate{R^{(1,2)}(\theta_1)} \vqw{2} 
& \gateO{2} \vqw{3} & \qw \\
\vdots \\
\lstick{$\ell-1$} & \qw  & \gateO{2} \vqw{1}  & \qw & \qw  \\
\lstick{$\ell=m-1$} & \gate{X^{(1,2)}}  &  \gateO{2}  
& \gate{X^{(1,2)}}  & \qw 
\end{quantikz}
\end{adjustbox}
\caption{Circuit diagram for $T^{(1)}_{m,k}= T^{(1)}_{m,2\ell+1} = T^{(1)\, 1}_{m; \ell}$ ($k$ odd), with $\ell = m-1, m>2$}
\label{fig:T1oddb}
\end{figure}
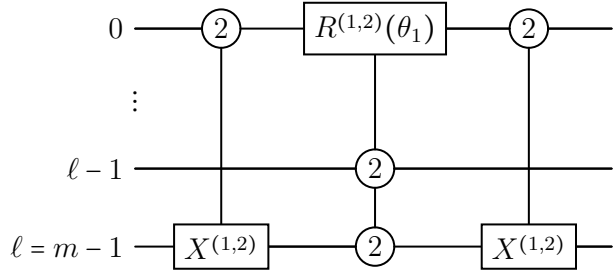

 \begin{figure}[htb]
	\centering
\begin{adjustbox}{width=0.4\textwidth}
\begin{quantikz}
\lstick{$0$} & \gateO{2} \vqw{1} & \gate{R^{(1,2)}(\theta_1)} \vqw{1} 
& \gateO{2} \vqw{1} & \qw \\
\lstick{$\ell=1$} & \gate{X^{(1,2)}}  &  \gateO{2}  
& \gate{X^{(1,2)}}  & \qw 
\end{quantikz}
\end{adjustbox}
\caption{Circuit diagram for $T^{(1)}_{m,k}= T^{(1)}_{m,2\ell+1} = T^{(1)\, 1}_{m; \ell}$ ($k$ odd), with $\ell=1, m=2$}
\label{fig:T1oddc}
\end{figure}
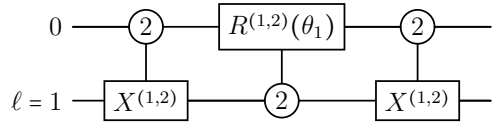

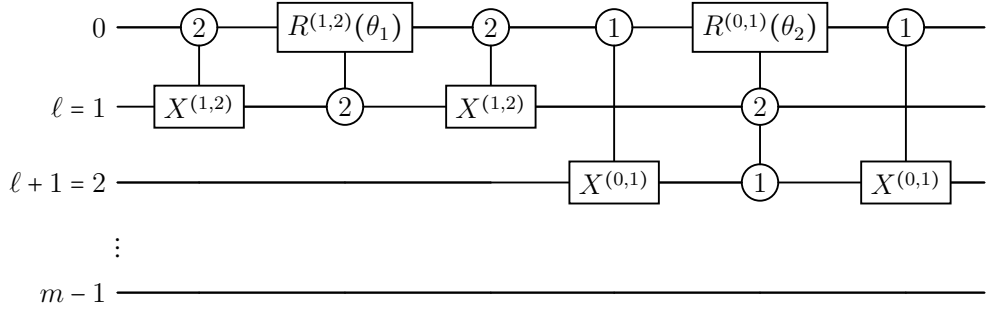
\begin{figure}[htb]
	\centering
\begin{adjustbox}{width=0.8\textwidth}
\begin{quantikz}
\lstick{$0$} & \gateO{2} \vqw{1} & \gate{R^{(1,2)}(\theta_1)} \vqw{1} 
& \gateO{2} \vqw{1}   & \gateO{1} \vqw{2} 
& \gate{R^{(0,1)}(\theta_2)} \vqw{1} & \gateO{1} \vqw{2}  & \qw \\
\lstick{$\ell=1$} & \gate{X^{(1,2)}}  &  \gateO{2}  
& \gate{X^{(1,2)}} & \qw  & \gateO{2} \vqw{1}
& \qw & \qw \\
\lstick{$\ell+1=2$} & \qw  &\qw  &\qw  & \gate{X^{(0,1)}} &\gateO{1} 
& \gate{X^{(0,1)}}  & \qw \\
\vdots\\
\lstick{$m-1$}&\qw  &\qw  &\qw  &\qw &\qw &\qw &\qw 
\end{quantikz}
\end{adjustbox}
\caption{Circuit diagram for $T^{(1)}_{m,k}= T^{(1)}_{m,2\ell+1} = T^{(1)\, 1}_{m; \ell}$ ($k$ odd), with $\ell=1, m>2$}
\label{fig:T1oddd}
\end{figure}

\begin{figure}[htb]
	\centering
\begin{adjustbox}{width=0.4\textwidth}
\begin{quantikz}
\lstick{$\ell=0$} & \gateO{1} \vqw{1} 
& \gate{R^{(0,1)}(\theta_2)} \vqw{1} & \gateO{1} \vqw{1}  & \qw \\
\lstick{$\ell+1=1$} & \gate{X^{(0,1)}} &\gateO{1} 
& \gate{X^{(0,1)}}  & \qw \\
\vdots\\
\lstick{$m-1$} &\qw &\qw &\qw &\qw 
\end{quantikz}
\end{adjustbox}
\caption{Circuit diagram for $T^{(1)}_{m,k}= T^{(1)}_{m,2\ell+1} = T^{(1)\, 1}_{m; \ell}$ ($k$ odd), with $\ell=0, m>1$}
\label{fig:T1odde}
\end{figure}
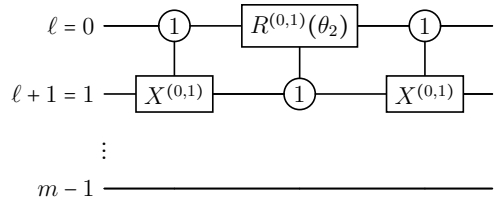

\clearpage

\begin{thebibliography}{10}

\bibitem{Nielsen:2019}
M.~A. Nielsen and I.~L. Chuang, {\em {Quantum computation and quantum
  information}}.
\newblock Cambridge University Press, 2019.

\bibitem{Barenco:1995}
A.~{Barenco}, C.~H. {Bennett}, R.~{Cleve}, D.~P. {Divincenzo}, N.~{Margolus},
  P.~{Shor}, T.~{Sleator}, J.~A. {Smolin}, and H.~{Weinfurter}, ``{Elementary
  gates for quantum computation},''
  \href{http://dx.doi.org/10.1103/PhysRevA.52.3457}{{\em Phys. Rev. A}
  {\bfseries 52} no.~5, (Nov., 1995) 3457--3467},
  \href{http://arxiv.org/abs/quant-ph/9503016}{{\ttfamily
  arXiv:quant-ph/9503016 [quant-ph]}}.

\bibitem{Kaye:2004}
P.~{Kaye} and M.~{Mosca}, ``{Quantum Networks for Generating Arbitrary Quantum
  States},'' in {\em {Optical Fiber Communication Conference and International
  Conference on Quantum Information ICQI}}, p.~PB28.
\newblock 2004.
\newblock \href{http://arxiv.org/abs/quant-ph/0407102}{{\ttfamily
  arXiv:quant-ph/0407102 [quant-ph]}}.

\bibitem{Mottonen:2004}
M.~{Mottonen}, J.~J. {Vartiainen}, V.~{Bergholm}, and M.~M. {Salomaa},
  ``{Transformation of quantum states using uniformly controlled rotations},''
  {\em Quant. Inf. Comp.} {\bfseries 5} (2005) 467,
  \href{http://arxiv.org/abs/quant-ph/0407010}{{\ttfamily
  arXiv:quant-ph/0407010 [quant-ph]}}.

\bibitem{Shende:2006}
V.~V. {Shende}, S.~S. {Bullock}, and I.~L. {Markov}, ``{Synthesis of
  quantum-logic circuits},'' {\em IEEE Trans on Computer-Aided Design}
  {\bfseries 25} (2006) 1000,
  \href{http://arxiv.org/abs/quant-ph/0406176}{{\ttfamily
  arXiv:quant-ph/0406176 [quant-ph]}}.

\bibitem{Plesch:2011}
M.~{Plesch} and {\v{C}}.~{Brukner}, ``{Quantum-state preparation with universal
  gate decompositions},''
  \href{http://dx.doi.org/10.1103/PhysRevA.83.032302}{{\em Phys. Rev. A}
  {\bfseries 83} no.~3, (2011) 032302},
  \href{http://arxiv.org/abs/1003.5760}{{\ttfamily arXiv:1003.5760
  [quant-ph]}}.

\bibitem{Dicke:1954zz}
R.~H. Dicke, ``{Coherence in Spontaneous Radiation Processes},'' {\em Phys.
  Rev.} {\bfseries 93} (1954) 99--110.

\bibitem{Prevedel:2009}
R.~{Prevedel}, G.~{Cronenberg}, M.~S. {Tame}, M.~{Paternostro}, P.~{Walther},
  M.~S. {Kim}, and A.~{Zeilinger}, ``{Experimental Realization of Dicke States
  of up to Six Qubits for Multiparty Quantum Networking},'' {\em Phys. Rev.
  Lett.} {\bfseries 103} no.~2, (2009) 020503,
  \href{http://arxiv.org/abs/0903.2212}{{\ttfamily arXiv:0903.2212
  [quant-ph]}}.

\bibitem{Toth:2012}
G.~{T{\'o}th}, ``{Multipartite entanglement and high-precision metrology},''
  {\em Phys. Rev. A} {\bfseries 85} no.~2, (2012) 022322,
  \href{http://arxiv.org/abs/1006.4368}{{\ttfamily arXiv:1006.4368
  [quant-ph]}}.

\bibitem{Farhi:2014}
E.~{Farhi}, J.~{Goldstone}, and S.~{Gutmann}, ``{A Quantum Approximate
  Optimization Algorithm},'' \href{http://arxiv.org/abs/1411.4028}{{\ttfamily
  arXiv:1411.4028 [quant-ph]}}.

\bibitem{Bartschi2019}
A.~B\"artschi and S.~Eidenbenz, ``{Deterministic preparation of Dicke
  states},'' {\em Lecture Notes in Computer Science} (2019) 126--139,
  \href{http://arxiv.org/abs/1904.07358}{{\ttfamily arXiv:1904.07358
  [quant-ph]}}.

\bibitem{Chakraborty:2012}
K.~{Chakraborty}, B.-S. {Choi}, A.~{Maitra}, and S.~{Maitra}, ``{Efficient
  quantum algorithm to construct arbitrary Dicke states},'' {\em Quant. Inf.
  Comp.} {\bfseries 13} no.~9, (2014) 2049--2069,
  \href{http://arxiv.org/abs/1209.5932}{{\ttfamily arXiv:1209.5932
  [quant-ph]}}.

\bibitem{Bartschi:2022}
A.~{B{\"a}rtschi} and S.~{Eidenbenz}, ``{Short-Depth Circuits for Dicke State
  Preparation},'' in {\em {2022 IEEE Int. Conf. Quant. Comp. Eng.}},
  pp.~87--96.
\newblock 2022.
\newblock \href{http://arxiv.org/abs/2207.09998}{{\ttfamily arXiv:2207.09998
  [quant-ph]}}.

\bibitem{VanDyke:2021kvq}
J.~S. Van~Dyke, G.~S. Barron, N.~J. Mayhall, E.~Barnes, and S.~E. Economou,
  ``{Preparing Bethe Ansatz Eigenstates on a Quantum Computer},'' {\em PRX
  Quantum} {\bfseries 2} (2021) 040329,
  \href{http://arxiv.org/abs/2103.13388}{{\ttfamily arXiv:2103.13388
  [quant-ph]}}.

\bibitem{VanDyke:2021nuz}
J.~S. Van~Dyke, E.~Barnes, S.~E. Economou, and R.~I. Nepomechie, ``{Preparing
  exact eigenstates of the open XXZ chain on a quantum computer},'' {\em J.
  Phys. A} {\bfseries 55} no.~5, (2022) 055301,
  \href{http://arxiv.org/abs/2109.05607}{{\ttfamily arXiv:2109.05607
  [quant-ph]}}.

\bibitem{Li:2022czv}
W.~Li, M.~Okyay, and R.~I. Nepomechie, ``{Bethe states on a quantum computer:
  success probability and correlation functions},'' {\em J. Phys. A} {\bfseries
  55} no.~35, (2022) 355305, \href{http://arxiv.org/abs/2201.03021}{{\ttfamily
  arXiv:2201.03021 [quant-ph]}}.

\bibitem{Bethe:1931hc}
H.~Bethe, ``{On the theory of metals. 1. Eigenvalues and eigenfunctions for the
  linear atomic chain},'' {\em Z. Phys.} {\bfseries 71} (1931) 205--226.

\bibitem{Gaudin:1983}
M.~Gaudin, {\em {La fonction d'onde de Bethe}}.
\newblock Masson, 1983.
\newblock English translation by J.-S. Caux, {\em The Bethe wavefunction}, CUP,
  2014.

\bibitem{Nepomechie:2023lge}
R.~I. Nepomechie and D.~Raveh, ``{Qudit Dicke state preparation},'' {\em
  Quantum Inf. Comp.} {\bfseries 24} (2024) 0037--0056,
  \href{http://arxiv.org/abs/2301.04989}{{\ttfamily arXiv:2301.04989
  [quant-ph]}}.

\bibitem{Li:2015}
Z.-H. {Li} and A.-M. {Wang}, ``{Entanglement entropy in quasi-symmetric
  multi-qubit states},'' {\em Int. J. Quant. Inf.} {\bfseries 13} no.~02,
  (2015) 1550007, \href{http://arxiv.org/abs/1310.3089}{{\ttfamily
  arXiv:1310.3089 [quant-ph]}}.

\bibitem{Raveh:2023iyy}
D.~Raveh and R.~I. Nepomechie, ``{$q$-analog qudit Dicke states},'' {\em J.
  Phys. A} (2024) , \href{http://arxiv.org/abs/2308.08392}{{\ttfamily
  arXiv:2308.08392 [quant-ph]}}.

\bibitem{Mozafari:2022}
F.~{Mozafari}, G.~{De Micheli}, and Y.~{Yang}, ``{Efficient deterministic
  preparation of quantum states using decision diagrams},''
  \href{http://dx.doi.org/10.1103/PhysRevA.106.022617}{{\em Phys. Rev. A}
  {\bfseries 106} no.~2, (Aug., 2022) 022617},
  \href{http://arxiv.org/abs/2206.08588}{{\ttfamily arXiv:2206.08588
  [quant-ph]}}.

\bibitem{Yeh:2023}
L.~{Yeh}, ``{Scaling W state circuits in the qudit Clifford hierarchy},'' in
  {\em {Proceedings of QP2023}}.
\newblock 2023.
\newblock \href{http://arxiv.org/abs/2304.12504}{{\ttfamily arXiv:2304.12504
  [quant-ph]}}.

\bibitem{Crampe:2011}
N.~{Cramp{\'e}}, E.~{Ragoucy}, and L.~{Alonzi}, ``{Coordinate Bethe Ansatz for
  Spin s XXX Model},'' {\em SIGMA} {\bfseries 7} (2011) 006,
  \href{http://arxiv.org/abs/1009.0408}{{\ttfamily arXiv:1009.0408 [math-ph]}}.

\bibitem{Lipkin:1964yk}
H.~J. Lipkin, N.~Meshkov, and A.~J. Glick, ``{Validity of many-body
  approximation methods for a solvable model. 1. Exact solutions and
  perturbation theory},''
  \href{http://dx.doi.org/10.1016/0029-5582(65)90862-X}{{\em Nucl. Phys.}
  {\bfseries 62} (1965) 188--198}.

\bibitem{Liu:2015}
W.-F. Liu and Z.-D. Hu, ``{Constructions of Dicke states in high spin
  multi-particle systems},'' \href{http://arxiv.org/abs/1511.03281}{{\ttfamily
  arXiv:1511.03281 [quant-ph]}}.

\bibitem{cirq}
{Quantum AI team and collaborators}, ``qsim,'' Sep, 2020.
\newblock \url{https://doi.org/10.5281/zenodo.4023103}.

\bibitem{Di:2011}
Y.-M. {Di} and H.-R. {Wei}, ``{Synthesis of multivalued quantum logic circuits
  by elementary gates},'' {\em Phys. Rev. A} {\bfseries 87} (2013) 012325,
  \href{http://arxiv.org/abs/1105.5485}{{\ttfamily arXiv:1105.5485
  [quant-ph]}}.

\bibitem{Wang:2020}
Y.~{Wang}, Z.~{Hu}, B.~C. {Sanders}, and S.~{Kais}, ``{Qudits and
  high-dimensional quantum computing},'' {\em Front. Phys.} {\bfseries 8}
  (2020) 479, \href{http://arxiv.org/abs/2008.00959}{{\ttfamily
  arXiv:2008.00959 [quant-ph]}}.

\bibitem{Wang:2021}
Y.~Wang and B.~M. Terhal, ``{Preparing Dicke states in a spin ensemble using
  phase estimation},'' {\em Phys. Rev. A} {\bfseries 104} no.~3, (2021) ,
  \href{http://arxiv.org/abs/2104.14310}{{\ttfamily arXiv:2104.14310
  [quant-ph]}}.

\bibitem{Piroli:2024ckr}
L.~Piroli, G.~Styliaris, and J.~I. Cirac, ``{Approximating many-body quantum
  states with quantum circuits and measurements},''
  \href{http://arxiv.org/abs/2403.07604}{{\ttfamily arXiv:2403.07604
  [quant-ph]}}.

\bibitem{Stanley2011}
R.~P. Stanley, {\em Enumerative Combinatorics}, vol.~1.
\newblock Cambridge University Press, 2011.
\newblock 2nd edition.

\bibitem{Popkov:2004}
V.~{Popkov} and M.~{Salerno}, ``{Logarithmic divergence of the block
  entanglement entropy for the ferromagnetic Heisenberg model},''
  \href{http://dx.doi.org/10.1103/PhysRevA.71.012301}{{\em Phys. Rev. A}
  {\bfseries 71} no.~1, (Jan., 2005) 012301},
  \href{http://arxiv.org/abs/quant-ph/0404026}{{\ttfamily
  arXiv:quant-ph/0404026 [quant-ph]}}.

\bibitem{Latorre:2004qn}
J.~I. Latorre, R.~Orus, E.~Rico, and J.~Vidal, ``{Entanglement entropy in the
  Lipkin-Meshkov-Glick model},''
  \href{http://dx.doi.org/10.1103/PhysRevA.71.064101}{{\em Phys. Rev. A}
  {\bfseries 71} (2005) 064101},
  \href{http://arxiv.org/abs/cond-mat/0409611}{{\ttfamily
  arXiv:cond-mat/0409611}}.

\bibitem{Lucke:2014mcw}
B.~L\"ucke, J.~Peise, G.~Vitagliano, J.~Arlt, L.~Santos, G.~T\'oth, and
  C.~Klempt, ``{Detecting Multiparticle Entanglement of Dicke States},''
  \href{http://dx.doi.org/10.1103/PhysRevLett.112.155304}{{\em Phys. Rev.
  Lett.} {\bfseries 112} no.~15, (2014) 155304}.

\bibitem{Moreno:2018}
M.~G.~M. {Moreno} and F.~{Parisio}, ``{All bipartitions of arbitrary Dicke
  states},'' \href{http://arxiv.org/abs/1801.00762}{{\ttfamily arXiv:1801.00762
  [quant-ph]}}.

\bibitem{Munizzi:2023ihc}
W.~Munizzi and H.~J. Schnitzer, ``{Entropy cones and entanglement evolution for
  Dicke states},'' \href{http://dx.doi.org/10.1103/PhysRevA.109.012405}{{\em
  Phys. Rev. A} {\bfseries 109} no.~1, (2024) 012405},
  \href{http://arxiv.org/abs/2306.13146}{{\ttfamily arXiv:2306.13146
  [quant-ph]}}.

\bibitem{Popkov:2005}
V.~{Popkov}, M.~{Salerno}, and G.~{Sch{\"u}tz}, ``{Entangling power of
  permutation-invariant quantum states},''
  \href{http://dx.doi.org/10.1103/PhysRevA.72.032327}{{\em Phys. Rev. A}
  {\bfseries 72} no.~3, (Sept., 2005) 032327},
  \href{http://arxiv.org/abs/quant-ph/0506209}{{\ttfamily
  arXiv:quant-ph/0506209 [quant-ph]}}.

\bibitem{Carrasco:2015sxh}
J.~A. Carrasco, F.~Finkel, A.~Gonz\'alez-L\'opez, M.~A. Rodr\'\i{}guez, and
  P.~Tempesta, ``{Generalized isotropic
  Lipkin\textendash{}Meshkov\textendash{}Glick models: ground state
  entanglement and quantum entropies},''
  \href{http://dx.doi.org/10.1088/1742-5468/2016/03/033114}{{\em J. Stat.
  Mech.} {\bfseries 1603} no.~3, (2016) 033114},
  \href{http://arxiv.org/abs/1511.09346}{{\ttfamily arXiv:1511.09346
  [quant-ph]}}.

\end{thebibliography}
\providecommand{\href}[2]{#2}\begingroup\raggedright\endgroup

\end{document}